\documentclass[twoside]{article}
\usepackage{fleqn}
\usepackage{espcrc2}
\usepackage{graphicx}
\usepackage{epsfig}
\usepackage{amsmath}
\usepackage[figuresright]{rotating}

\newcommand{\be}{\begin{equation}}
\newcommand{\ee}{\end{equation}}
\newcommand{\bea}{\begin{eqnarray*}}
\newcommand{\eea}{\end{eqnarray*}}
\newcommand{\bqa}{\begin{eqnarray*}}
\newcommand{\eqa}{\end{eqnarray*}}

\newcommand{\nl}{\nonumber \\}
\newcommand{\ep}{\varepsilon}
\newcommand{\eps}{\varepsilon}

\begin{document}
\title{%
{\normalsize \tt
DESY 06-010
\\
WUE-ITP-2006-001
\\
SFB/CPP-06-03 
\\
}
\vspace*{.5cm}
Differential equations and massive two-loop Bhabha scattering: the B5l2m3 case%
\thanks{%
Contribution to RADCOR,
Oct 2-7, 2005, Shonan Village, Japan, to appear in the proceedings.}%
\thanks{Work supported in part by
Sonderforschungsbereich/Transregio 9--03 of DFG
`Computergest{\"u}tzte Theo\-re\-ti\-sche Teil\-chen\-phy\-sik',  by
the Sofja Kovalevskaja Award of the Alexander von Humboldt Foundation
  sponsored by the German Federal Ministry of Education and Research,
and by the Polish State Committee for Scientific Research (KBN),
research projects in 2004--2005.
}
}
\author{M. Czakon%
\address{Institut f\"ur Theoretische Physik
und Astrophysik, Universit\"at W\"urzburg,
Am Hubland, D-97074 W\"urzburg, Germany}%
,
J. Gluza$^{\rm b}$,
K. Kajda\address{Institute of Physics, University of Silesia, 
   ul. Uniwersytecka 4, 40007 Katowice, Poland}
        ~and
        T. Riemann\address{Deutsches Elektronen-Synchrotron, DESY, 
   Platanenallee 6, 15738 Zeuthen, Germany}
}
\begin{abstract}
The two-loop box contributions to massive Bhabha scattering may be reduced to two-loop box master integrals (MIs) with five, six, and seven internal lines, plus vertices and self energies.
The self-energy and  vertex MIs may be solved analytically by the
differential equations (DE) method.
This is true for only few of the box masters.
Here we describe some details of the analytical determination, including constant terms in $\ep = (4 - d) /2$, of the complicated topology B5l2m3 (with 5 lines, 2 of them being massive).
With the  DE approach, three of the four coupled masters have been solved
in terms of (generalized) standard Harmonic Polylogarithms.
\end{abstract}
\maketitle

\section{\label{sec-intro}INTRODUCTION}
Bhabha scattering,
\begin{eqnarray*}
e^+e^- \to e^+e^-(\gamma),
\end{eqnarray*}
might be used as an extremely accurate luminosity monitor at the planned International Linear Collider ILC, with an accuracy of up to $10^{-4}$ in the very forward direction.
Some more details on the general status of this topic may be found in \cite{Czakon:2005gi,Czakon:2005radc,Bonciani:2006qu} and in references therein.
To meet these accuracy needs by theory, one has to control the virtual two-loop corrections.
This has been substantially pushed forward quite recently in \cite{Penin:2005kf,Penin:2005eh} by a derivation of the constant terms of the cross section expansion in the small parameters $m^2/s, m^2/t$ from a massless calculation  \cite{Bern:2000ie}, combined with contributions from closed fermion loop insertions \cite{Bonciani:2005im}.
Alternatively, we are performing a direct and independent calculation of the massive two-loop Feynman integrals.
Our method relies on the calculation of a relatively small number of scalar master integrals and the use of algebraic relations for expressing all the others.
The two-loop self energy and vertex master integrals are known from
\cite{Bonciani:2003te,Bonciani:2003hc,Czakon:2004tg,Czakon:2004wm} and the references therein; for additional informations and an introduction to our conventions, see also the webpage \cite{Czakon:2004n1}.
For their determination, the method of differential equations has been used \cite{Kotikov:1991hm,Laporta:1996mq,Remiddi:1997ny,Laporta:2001dd}.
The
iterative, nested integration of a system of DEs for MIs of rising complexity was possible by use of a certain class of functions, the so-called Harmonic Polylogarithms (HPLs) \cite{Remiddi:1999ew} and Generalized Harmonic Polylogarithms (GPLs) \cite{Gehrmann:2000zt}.
This works quite nicely for problems with one scale (with HPLs) and for some problems with two scales (GPLs); see the references mentioned.
For two-loop Bhabha scattering, the question arises whether a full analytical solution can be obtained with HPLs
with arguments $(+1,0,-1)$ and GPLs with arguments $(+1,0,-1,y,-1/y)$.
The answer is positive for MIs for self-energies (SE) and vertices.
Presently, we are determining all the box masters needed.
For the MIs of box type we know a few solutions in terms of HPLs and GPLs.
However, most cases have not been solved so far.
One reason is the appearance of systems of coupled DEs.
B6l3m3 e.g. is a box topology with 6 lines, 3 of them massive, and one is faced with a system of six coupled DEs.
For the topologies B5l3m and B5l2m3, there are five and four coupled master equations, respectively.

Here, we present analytical results for topology B5l2m3, which have been obtained as solutions of differential equations.
B5l2m3 appears to be a complicated and very interesting case.
This topology is part of
decompositions for the topologies B1 and B3 \cite{Czakon:2004wm}.
The basic master B7l4m1 for the topology B1 is known \cite{Smirnov:2001cm}, and expressed through HPLs and GPLs.
So, it is natural to assume that its subdiagrams, including B5l2m3
should have a similar structure.
However, the situation might be different.

\section{\label{sec-topo}TWO-LOOP BOX TOPOLOGIES AND MASTER INTEGRALS}
There are six double box topologies \cite{Czakon:2004wm}.
Three of them have master topologies with up to seven lines: the planar B1, the second planar B2, and the nonplanar B3.
For a reduction, one needs also master integrals with six and five internal lines. Only few of them are known analytically.
The completely known double boxes with five lines are B5l4m \cite{Bonciani:2003cj,Czakon:2004tg} and B5l2m1 \cite{Czakon:2004tg}.
Further, the divergent parts of the B5l2m2 and B5l2m3 masters were determined in \cite{Czakon:2004wm} and those of B5l3m in \cite{Czakon:2005gi}.
Here, we will determine the finite parts of the three dotted masters for B5l2m3.
\section{\label{sec-b5l2m2} TOPOLOGY B5l2m3}
\begin{figure}[thb]
\begin{center}
 \epsfig{file=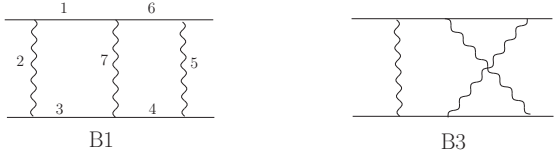, width=7.5cm}
\caption{\label{b1b3}The topologies B1 (planar) and B3 (nonplanar).}
\end{center}
\end{figure}

\begin{figure}[thb]
\begin{center}
 \epsfig{file=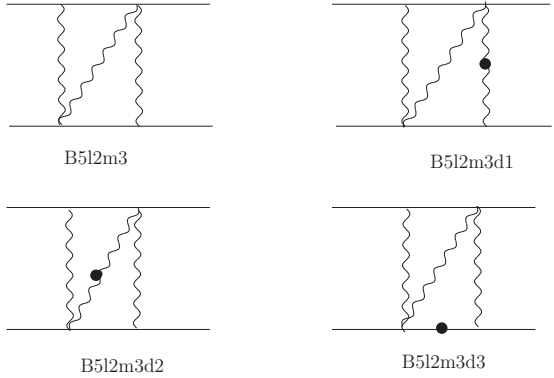, width=7.5cm}
\caption{\label{fig-b5l2m3}The masters for the topology B5l2m3.}
\end{center}
\end{figure}

The topology {\rm B5l2m3} contributes to the reduction of B1 and B3, see Figure \ref{b1b3}.
It may be obtained from B1 by shrinking lines 3 and 6,
\bea
\label{bh-48}
\nonumber
{\rm B5l2m3}[b_i] &=& \frac{e^{2\ep\gamma_E}}{\left(i\pi^{d/2}\right)^2} \int \frac{d^dk_1 d^dk_2 ~(k_2p_3)^{-b_8}}
     {D_1^{b_1} D_2^{b_2} D_4^{b_4} D_5^{b_5}D_7^{b_7}}  ,
\eea
where the arguments are $\{b_1~ b_2~ b_4~ b_5~ b_7~b_8\}$.
If not stated otherwise $p_4$ is eliminated by momentum conservation,
$p_1 + p_2 = p_3 + p_4$.
Further,
\bea
D_1 &=&(k_1+k_2+p_3)^2-1,
\\
D_2&=&(k_1+k_2-p_1+p_3)^2,
\\
D_4&=&(k_2-p_1-p_2+p_3)^2-1,
\\
D_5&=&k_2^2,
\\               
D_7&=&k_1^2
.
\eea
There are four masters, and one may choose those shown in Figure \ref{fig-b5l2m3}:
\bea
{\rm B5l2m3}   &=& {\rm B5l2m3}[111110],
\\
{\rm B5l2m3d1} &=& {\rm B5l2m3}[111210],
\\
{\rm B5l2m3d2} &=& {\rm B5l2m3}[111120],
\\
{\rm B5l2m3d3} &=& {\rm B5l2m3}[112110].
\eea

The singularities of the three dotted masters in $\ep = (4-d)/2$ have been determined recently using the method of differential equations \cite{Czakon:2004wm}.
For completeness, we quote them here:
\begin{eqnarray*}
\label{beta6d}
{\tt B5l2m3d1} = -\frac{1}{\epsilon^2} \frac{ (1 + x^2) y}{8 x (1-y)^2}
\nl
+~ \frac{1}{\epsilon}
\Bigl\{
 \frac{y(1 + x^2)}{4(1 - y)^2}
-\frac{2y^2(1-x)^2}{x(1 - y)(1 + y)^3} \zeta_2
\nl
 -~\frac{y}{2x(1 - y)^2(1 + y)^2} [ 2 y(1 + x^2)+ x (1 - y)^2]
\nl
  \left(H[0,y] +2 H[1,y] \right)
\nl
+~\frac{y(1 - x^2)}{4x(1 - y)^2}  H[0,x]
\nl
-\frac{y^2(1 - x)^2}{2x(1 - y)(1 + y)^3} \left(H[0,0,y] +2 H[0,1,y]\right)
\Bigr\}
\nl
+ {\cal{O}}(1),
\nonumber
\end{eqnarray*}

\begin{eqnarray*}
\label{beta6k}
{\tt B5l2m3d2}
=
-\frac{1}{\epsilon^2} \frac{x y H[0, x]}{(1 - x^2) (1 -
  y)^2}
\nl
-~
\frac{1}{\epsilon} \frac{2x y}{(1- x^2) (1 - y)^2}
(
H[1,0,x] -H[-1,0,x]
\nl
 + H[0,0,x]
+~\frac{\zeta_2}{2} +
H[0,x]( H[0,y]
\nl
+2~H[1,y])
)
+ {\cal{O}}(1),
\end{eqnarray*}

\begin{eqnarray*}\label{beta6l}
{\tt  B5l2m3d3} = -\frac{1}{\epsilon}  \frac{y}{2 (1 - y^2)} [ 4
  \zeta_2 +    H[0, 0, y]
\nl
  +  2   H[0, 1, y] ]
  + {\cal{O}}(1),
\nonumber
\end{eqnarray*}
and {\rm B5l2m3} is finite.
The functions $H$ are Harmonic Polylogarithms (HPLs) in the variables
$x = (\sqrt{-s+4} - \sqrt{-s})/ (\sqrt{-s+4} + \sqrt{-s})$ and $y$ (obtained by replacing $s$ by $t$).
A basis for use in Mathematica may be found in file {\tt HPL4.m} at \cite{Czakon:2004n1}.

Three of the finite parts of the masters were found in several steps.
Only B5l2m3d1 and B5l2m3d2 have terms of order $O(1/\ep^2)$.
The B5l2m3d2 decouples in a basis where B5l2m3d1 is replaced: \{B5l2m3, B5l2m3d2,
B5l2m3d3, B5l2m3N\}.
The  B5l2m3N = \mbox{B5l2m3[11111-1]} with irreducible numerator is of order $O(1)$ in $\eps$.
In fact, there are four irreducible numerators.
For the momenta chosen here, they are either:
($k_1p_2, k_1p_3, k_2p_3$) and ($k_1p_1$ or $k_2p_1$ or $k_2p_2$), or alternatively:
($k_1k_2, k_1p_2$) and ($k_1p_3$ or $k_2p_3$) and ($k_2p_1$ or $k_2p_2$).
Any of the numerator integrals may be used to replace a dotted master.
The corresponding algebraic relation for the chosen case relates
{\tt B5l2m3N} with {\tt B5l2m3}, {\tt B5l2m3d1}, {\tt B5l2m3d2}, {\tt B5l2m3d3}, ${\tt T1l1m2}^2$, {\tt SE3l2m}(s),
{\tt SE3l1mONS}, {\tt SE3l0m}(t), {\tt V4l2m2}(s), {\tt V4l1m2}(t), and
will be reproduced in \cite{Czakon:2004n1}.
In this basis we may derive for ${\tt B5l2m3d2[0,x,y]}$ a decoupled first-order $s$-channel DE, which can be easily solved with a boundary condition at $y = 1$:
\begin{eqnarray*}
\label{beta6d20}
 {\tt B5l2m3d2}[0,x,y]
 =
 \frac{-4xy}{(1 - x^2)(1 - y)^2}
 \nl
\Bigl\{
        \frac{\zeta_2}{4}
        ( 20H[ - 1/y,x] - 12H[ - y,x]
        - 10 H[-1,x]
\nl
        - 3 H[0,x] - 14
         H[0,y]
+ 2 H[1,x] + 4 H[1,y] )
\nl
 +   7/4 ~\zeta_3
\nl
 - 3H[-1,0,x]( H[0,y] +2 H[1,y])
\nl
  + 2H[0,x] ( H[1,0,y]  + 2 H[1,1,y])
\nl
   + (H[0,y]+2H[1,y])
(H[-1/y,0,x]
\nl
+ H[ - y,0,x] + H[1,0,x] + H[0,0,x])
\nl
+ H[0,0,y](H[ - 1/y,x]
-H[ - y,x] + H[0,x])
\nl
+ H[1,1,0,x] - H[0,0,0,y] - 2H[0,0,1,y]
\nl
+ 2H[0,1,y](H[ - 1/y,x] - H[ - y,x] + 2H[0,x])
\nl
       + 2H[ - 1/y,-1,0,x] - H[ - 1/y,0,0,x]
\nl
    + 2H[ - y,-1,0,x]
       - H[ - y,0,0,x]
\nl
  + H[ -1,0,0,x]
  - H[-1,1,0,x]
- H[0,-1,0,x]
  \nl
 + H[0,1,0,x]
 - H[1,-1,0,x]
 + H[1,0,0,x]
 \nl
-3H[-1,-1,0,x]\Bigr\}
.
\nonumber
\end{eqnarray*}
In a next step B5l2m3d3 can be derived using the same basis.
However, in the differential equation the homogeneous part is absent.
This means that we are not able to formulate a boundary condition for this equation using the analyticity of the solution.
However, an analogous situation appears for the differential equation for
${\tt B5l2m3d3[0,x,y]}$ in the $t$-channel.
So, we have two solutions (which must give finally the same numbers), but up to the `constant' terms $c_1(y)$ in the $s$-channel and $c_2(x)$ in the $t$-channel.
We take $x=1$ and get in the $s$-channel a pure function of $y$.
Comparing now the two solutions we may determine $c_1(y)$ up to an unknown true constant $c_2(1)$.
For HPLs of weight three, this constant is proportional to $\zeta_3$ and can
be fitted numerically:
\begin{eqnarray*}
\label{beta6d3}
{\tt B5l2m3d3}[0,x,y]
 = \frac{y}{1-y^2}
\{
\zeta_2(4H[-1, y]
\nl
- 7/2H[0, y]
- 8H[1, y]
-5H[-1/y, x]
\nl
- 3H[-y, x]+4 H[0,x])
\nl
  -~\zeta_3/2
  \nl
+H[-1, 0, 0, y]
+ 2H[-1, 0, 1, y]
\nl
 - 5/2H[0, 0, 0, y] - 5H[0, 0, 1, y]
-  2H[0, 1, 0, y]
\nl
 - 4H[0, 1, 1, y]  + H[-1/y, 0, 0, x]
\nl
  - 4H[1, 0, 1, y] -
  2H[-1/y, -1, 0, x]
\nl
  - 2H[1, 0, 0, y] + 2H[-y, -1, 0, x] -
  H[-y, 0, 0, x]
\nl
  -(H[-1/y, 0, x]\!- \!\! H[-y, 0, x])(H[0, y]+2H[1, y])
\nl
-(H[-1/y, x]
+H[-y, x]-H[0, x])
\nl
(H[0, 0, y]+2H[0, 1, y])
\}
.
\nonumber
\end{eqnarray*}
Finally,
the solution for B5l2m3d1 can be found with the original basis \{B5l2m3, B5l2m3d1, B5l2m3d2, B5l2m3d3\} and using the fact that B5l2m3 is not singular in $\ep$ (this can be seen by IR power counting, but has also been checked with a sector decomposition).
The vanishing of the singularity of B5l2m3 yields an algebraic relation between
${\tt B5l2m3d1}[0,x,y]$  and ${\tt B5l2m3d3}[0,x,y]$ which may be easily resolved for
${\tt B5l2m3d1}[0,x,y]$:
\allowdisplaybreaks
\begin{eqnarray*}
\label{beta6d1}
{\tt B5l2m3d1}[0,x,y]=
\frac{-y(1-x^2)}{2x(1 - y^2)^2}
\{ (3-2y+3y^2)
\nl
H[-1,0,x]
-4y
H[0,x](H[0,y]+2H[1,y])
\nl
- (1+y)^2(H[1,0,x]-H[0,x])
\}
\nl
+\frac{y}{8x(1 - y)^2(1 + y)^2}
\{8(2y(1+x^2)+x(1-y)^2)
\nl
(H[0,y]-2H[1,0,y]
+2H[1,y] -4H[1,1,y])
\nl
-4(1-2x+5x^2+2y+4xy+2x^2y+y^2
\nl
-2xy^2+5x^2y^2)
H[0,0,x]
\nl
-4(3x+4y-4xy+4x^2y+xy^2)
\nl
(H[0,0,y] +2H[0,1,y])
\nl
-(15+4x+11x^2-10y+24xy
\nl
+14x^2y+15y^2-28xy^2+11x^2y^2)\zeta_2
\}
\nl
-\frac{y(1+x^2)}{2x(1 - y)^2}
\nl
-\frac{(1-x)^2y^2}{x(1 - y)(1 + y)^3}
\{2H[-1/y, -1, 0, x]
\nl
-H[-1/y, 0, 0, x]
+H[-y, 0, 0, x]
\nl
-2H[-y, -1, 0, x] - H[-1, 0, 0, y]
\nl
+ 5H[0, 0, 1, y]
-2H[-1, 0, 1, y]
\nl
+ 5/2H[0, 0, 0, y]
+2H[0, 1, 0, y]
\nl
+4H[0, 1, 1, y] +2H[1, 0, 0, y]
\nl
+4H[1, 0, 1, y]
\nl
+\zeta_3/2
\nl
  - (4H[0,x] + 4H[-1,y] - 3H[-y,x]
\nl
 -5H[-1/y,x]- 7/2H[0,y]- 8H[1,y]) \zeta_2
\nl
 - (-H[-1/y,x]-H[-y,x] +H[0,x])
\nl
 (H[0,0,y] +2H[0,1,y])
\nl
  +(H[-1/y,0,x] -H[-y,0,x])
\nl
  (H[0,y]+2H[1,y])
  \} .
  \nonumber
\end{eqnarray*}
The final parts of the dotted masters depend on HPLs and additionally on generalized HPLs (GPLs) \cite{Gehrmann:2000zt}.
The latter are used here in the notations of \cite{Czakon:2004tg,Czakon:2004n1}.
A basis for use in Mathematica may be found in file {\tt GPL.m} at \cite{Czakon:2004n1}.
While ${\tt B5l2m3d2}[0,x,y]$ and ${\tt B5l2m3d3}[0,x,y]$ have a relatively simple algebraic structure in $x$ and $y$ and depend on functions with definite weights, both this is not true for ${\tt B5l2m3d1}[0,x,y]$: this function has a complicated algebraic structure and depends on HPLs and GPLs of mixed weights.
However, a comparison with ${\tt B5l2m3d1}[-1,x,y]$, the $1/\eps$-term in ${\tt B5l2m3d1}$, shows that both observations apply already there.

Finally, we are left with one unresolved term, namely the constant term of the basic master, ${\tt B5l2m3}[0,x,y]$.
For this, we have built many different bases of four MIs, with the aim to get finally a solvable differential equation for ${\tt B5l2m3}[0,x,y]$
with proper denominators (allowing us to integrate the DE with use of normal HPLs and GPLs).
However, so far we haven't found a basis with a proper structure.
It was possible to derive  two coupled differential equations
where ${\tt B5l2m3}[0,x,y]$ is involved, and then to get a second order differential equation for it.
We found that this equation has a structure which seems not to be integrable with normal HPLs and GPLs;
the ${\tt B5l2m3}[0,x,y]$ might reside in a different class of functions than used so far.

The masters described here will be posted in the Mathematica file {\tt MastersBhabha.m} at \cite{Czakon:2004n1}.
For the calculations we used {\tt MATHEMATICA}.
\section{\label{sec-sum}SUMMARY}
We have sketched here the derivation of analytical solutions for the constant parts of MIs  {\tt B5l2m3d1}[0,x,y], {\tt B5l2m3d2}[0,x,y], {\tt B5l2m3d3}[0,x,y].
The method of DEs was used.
 It is evident that for more complicated cases the procedure outlined here will not be applicable both for principal reasons and for numerical ones.
E.g., a fitting of constants as it was needed for B5l2m3d3 will in general not be  possible due to loss of precision of numeric calculations.
Further, when the number of coupled MIs is large ($\geq 3$), the situation starts to be complicated.
To find solutions,  an appropriate basis of MIs must be defined and differential equations must be
 prepared by applying differential operators.
Not always is the choice of MIs with dots and without numerators the best one; numerators may improve the infrared behavior and change the algebraic structure.
As mentioned, coupled systems of five or six MIs seem to be too complicated systems to be solved analytically.
We don't know which class of functions will be finally sufficient to cover all MIs for Bhabha scattering.
The topology B5l2m3 is the most complicated case so far, which has been - partly - solved analytically with the method of DE.

\end{document}